\documentclass[pre,twocolumn,showkeys,showpacs]{revtex4-1}

\usepackage{amsmath}    
\usepackage{bm}         
\usepackage{graphicx}   
\usepackage{color}      
\usepackage{subfigure}  
\usepackage{hyperref}   
\usepackage{bbold}

\begin{document}

\title{Compression- and Shear-Driven Jamming of U-Shaped Particles in Two Dimensions}
\author{Theodore A. Marschall}
\affiliation{
	Department of Physics and Astronomy, 
	University of Rochester, Rochester, NY 14627}
\author{Scott V. Franklin}
\affiliation{
	School of Physics and Astronomy, 
	Rochester Institute of Technology, Rochester, NY 14623 }
\author{S. Teitel}
\affiliation{Department of Physics and Astronomy, 
	University of Rochester, Rochester, NY 14627}


\begin{abstract}
We carry out numerical simulations of soft, U-shaped, frictionless particles in $d=2$ dimensions in order to explore the effects of complex particle shape on the jamming transition.  We consider both cases of uniform compression-driven and shear-driven jamming as packing fraction $\phi$ and compression or shear rate is varied.
Upon slow compression, jamming is found to occur when the isostatic condition is satisfied.  
Under driven steady state shearing, jamming occurs at a higher packing fraction $\phi_J$ than observed in compression.
A growing relaxation time and translational correlation length is found as $\phi$ increases towards $\phi_J$. 
We consider the orientational ordering and rotation of particles induced by the shear flow.
Both nematic and tetratic ordering are found, but these decrease as $\phi$ increases to $\phi_J$. At the jamming transition, the nematic ordering further decreases, while the tetratic ordering increases, but the orientational correlation lengths remain small throughout.  The average angular velocity of the particles is found to increase as $\phi$ increases, saturating to a plateau just below $\phi_J$, but then increasing again as $\phi$ increases above $\phi_J$. 
\vskip 12pt
\noindent [The final publication is available at Springer via http://dx.doi.org/10.1007/s10035-014-0540-2]

\end{abstract}

\maketitle

\section{Introduction}

Granular materials are found frequently in natural and industrial settings with a wide variety of different types of constituent particles.  
Considerable theoretical and numerical work has investigated the jamming transition in such granular systems \cite{Liu.Nature1998,VanHecke.CM2010,Torquato.RevPhys2010}, where the system undergoes a transformation from a liquid-like state to a rigid but disordered solid.
Most work has focused on the simplest case of spherical and circular particles \cite{OHern.PRE2003,Donev.JAP2004,Wyart.PRE2005,Somfai.PRE2007,Olsson.PRL2007}.  
Some recent works have considered packings of ellipsoidal particles \cite{Donev.PRL2004,Man.PRL2005,Donev.Science2004,Donev.PRE2007,Zeravcic.EPL2009,Mailman.PRL2009}, rods \cite{Narayan.JStatMech2006, Hidalgo, Desmond, Trepanier, Azema1, Azema2}, and polyhedra \cite{JiaoTorquato}.  Others have studied the behavior of rods and ellipsoids under shear driven flow \cite{Cleary,Campbell,Guo,Borzsonyi.PRL2012,Borzsonyi2,Guo.PhyFluids2013,Farhadi}. For a review of the effect of particle shape on granular properties, see Ref. \cite{Borzsonyi1}.
However it remains of interest to explore the effects that more complex particle shape may have on behavior in granular materials, and in particular on the jamming transition.

\begin{figure}
\centering
\includegraphics[width=0.75\columnwidth]{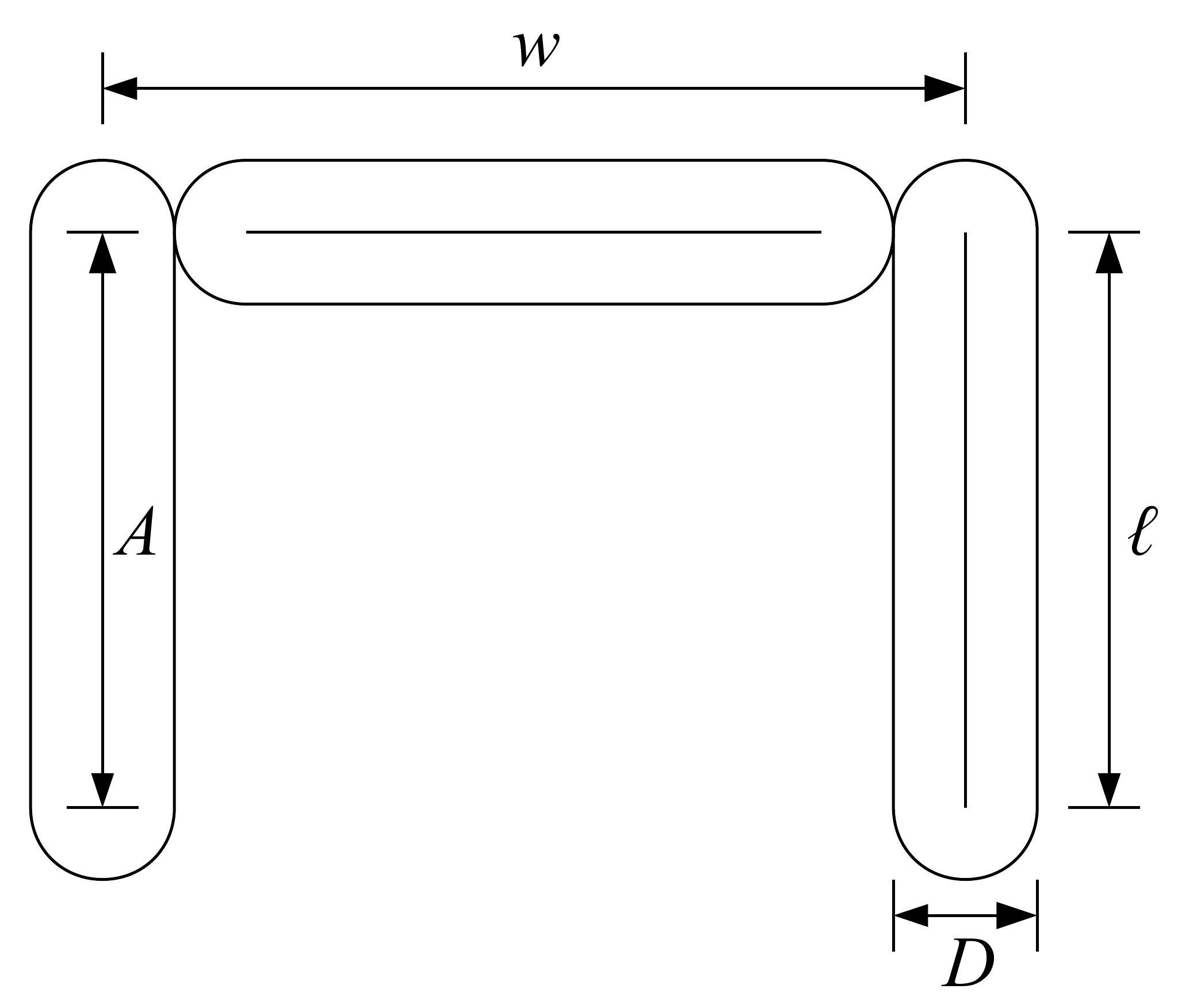}
\caption{\label{fig:staple}  The geometric model for a staple used in simulations consisting of three spherocylinders.  The spherocylnders are characterized by their diameters $D$ and axis lengths $A$, which determine the composite staple's spine length $w$ and barb length $\ell$.}
\end{figure}

Several recent works have explored the behavior of U-shaped particles, i.e. ``staples" (see Fig.~\ref{fig:staple}), which are interesting because their concave shape allows them to interlock, creating an effective inter-particle cohesion.  
In one study, this geometric cohesion was exhibited through the formation of free-standing columns of staples \cite{Gravish.PRL2012}.  
When allowed to collapse under vibration, the column height followed a stretched exponential, and the effective cohesion governing the rate of collapse was found to have a maximum value which depended on the spine-to-barb ratio of the staples.  
In another study, piles of U-shaped staples were subjected to extensional forces. The fluctuations in yield force were shown to result from the failure of weak links within the pile, well explained by a Weibullian weakest-link theory \cite{Franklin_rheology_2014}.

In this work we use numerical simulations of a two-dimensional system of frictionless, U-shaped, staples to explore whether such geometric cohesion has any significant effects on the jamming transition.  
We consider both uniform compression-driven jamming and uniform shear-driven jamming.  We investigate the response of the pressure in the system to different fixed compression and shear rates, and relate the onset of jamming to the isostatic condition \cite{Alexander.PhysRep1998} on the average particle contact number $\langle z\rangle$.  We also investigate the angular orientation and angular velocity of the staples in the sheared ensemble, which can be contrasted with the behavior of simpler elongated granular materials  \cite{Campbell,Guo,Borzsonyi.PRL2012, Borzsonyi2,Guo.PhyFluids2013,Farhadi}.

\section{Model}

Our model consists of a system of $N=1024$ identical frictionless staples in a two-dimensional, periodic, square box of side length $L$.  
Particles in contact interact with a repulsive elastic force.  
As contact detection for arbitrarily shaped particles is in general difficult, we model our staples as a rigid composite of three orthogonal spherocylinders, as illustrated in Fig.~\ref{fig:staple}.  
For spherocylinders, an efficient contact algorithm is known \cite{Pournin.GM2005}.  
The geometry of such staples are determined by the spherocylinders' axis length $A$ and diameter $D$ (see Fig.~\ref{fig:staple}).
All spherocylinders in this work are congruent, with an axis to diameter ratio $A/D = 4/1$, giving all of the resulting staples a spine length $w=A+2D$, barb length $\ell=A$, and barb to spine ratio of $\ell/w = 2/3$.
This is slightly higher than the ratio found for maximum cohesiveness in three dimensional columns \cite{Gravish.PRL2012}.  
We will assume that the staples have a total mass $m$ distributed uniformly over the area of the spherocylinders.  
The packing fraction $\phi$ of the system is given by,
\begin{equation}
\phi=N{\cal A}/L^2, \quad {\cal A}=3DA+3\pi (D/2)^2,
\end{equation}
with ${\cal A}$ the area of a single staple.

We define $\mathbf{r}_{ab}\equiv\mathbf{r}_b-\mathbf{r}_a$ as the shortest displacement between the axes of two spherocylinders $a$ and $b$, belonging to different staples $i$ and $j$, with $\mathbf{r}_a$ and $\mathbf{r}_b$ the corresponding points on the axes.  Two particles are considered to be in contact whenever this distance is less than (within machine precision) the particle diameter, $|\mathbf{r}_{ab}| <D$.
We then use a harmonic interaction for the elastic energy of two spherocylinders in contact,
\begin{equation}
U^\mathrm{el}(\mathbf{r}_{ab})=\left\{
\begin{array}{cr}
\frac{1}{2}k_e \left(1-|\mathbf{r}_{ab}|/D\right)^2, & |\mathbf{r}_{ab}|<D\>\\
0,&|\mathbf{r}_{ab}|\ge D,
\end{array}
\right.
\label{eq:Uel}
\end{equation}
with repulsive force $\mathbf{F}^\mathrm{el}_{ab}=-dU^\mathrm{el}/d\mathbf{r}_{a}$ acting upon spherocylinder $a$ at the point of contact $ab$.
This force is directed along the normal to the surface at the point of contact, pointing inwards to spherocylinder $a$.  
The total elastic force acting at the center of mass $\mathbf{r}_i$ of  staple $i$ is then the sum of all contact forces acting on its constituent spherocylinders,
\begin{equation}
\mathbf{F}^\mathrm{el}_i=\sum_{\mathrm{contacts }\, ab}\mathbf{F}^\mathrm{el}_{ab},
\label{eq:Fel}
\end{equation}
and the total torque about the staple's center of mass from these elastic forces is,
\begin{equation}
\tau^\mathrm{el}_i=\sum_{\mathrm{contacts }\, ab}\mathbf{\hat z}\cdot(\mathbf{r}_{iab}\times\mathbf{F}^\mathrm{el}_{ab}),
\label{eq:tauel}
\end{equation}
where $\mathbf{r}_{iab}$ is the displacement from the staple's center of mass $\mathbf{r}_i$ to the contact point $ab$, and $\mathbf{\hat z}$ is the unit normal perpendicular to the plane of the staples.

In addition to elastic contact forces, the staples also experience a viscous dissipative force.  
Following a commonly used simple model  \cite{Durian.PRL1995,Olsson.PRL2007}, we take this dissipative force to be proportional to the difference between the local velocity of each element of the staple and an average background velocity $\mathbf{v}_\mathrm{av}(\mathbf{r})$.  
We may think of this background $\mathbf{v}_\mathrm{av}(\mathbf{r})$ as representing either the average velocity of other staples at position $\mathbf{r}$, or as the velocity of a host fluid in which the staple is embedded.
If $\mathbf{v}_i\equiv\dot{ \mathbf{r}}_i$ is the velocity of the center of mass of staple $i$, and $\omega_i\equiv\dot\theta_i$ is its angular velocity about the center of mass, then the dissipative force per unit area acting at point $\mathbf{r}_i+\mathbf{r}^\prime$ on the staple (where $\mathbf{r}^\prime$ is the position relative to the center of mass $\mathbf{r}_i$) is,
\begin{equation}
\mathbf{f}_i^\mathrm{dis}(\mathbf{r}^\prime)=-k_d \left[\mathbf{v}_i+\omega_i\mathbf{\hat z}\times \mathbf{r}^\prime -\mathbf{v}_\mathrm{av}(\mathbf{r}_i+\mathbf{r}^\prime)\right].
\end{equation}
The total dissipative force acting at the staple's center of mass is then
\begin{equation}
\mathbf{F}^\mathrm{dis}_i=\int\limits_\mathrm{staple} d\mathbf{r}^\prime\>\mathbf{f}_i^\mathrm{dis}(\mathbf{r}^\prime)
\end{equation}
where $\mathbf{r}^\prime$ integrates over the area of the staple.  The total dissipative torque on the staple about its center of mass  is,
\begin{equation}
\tau_i^\mathrm{dis}=\int\limits_\mathrm{staple} d\mathbf{r}^\prime \mathbf{\hat z}\cdot[\mathbf{r}^\prime\times\mathbf{f}_i^\mathrm{dis}(\mathbf{r}^\prime)].
\end{equation}

To model a system uniformly compressed at a fixed rate $\epsilon$, we take as the average background velocity
\begin{equation}
\mathbf{v}_\mathrm{av}(\mathbf{r})=-\epsilon\mathbf{r}
\end{equation}
and use periodic boundary conditions on a box of length $L$ that shrinks at the same rate, $\dot L=-\epsilon L$.

Using $\int d\mathbf{r}^\prime={\cal A}$, $\int d\mathbf{r}^\prime \mathbf{r}^\prime = 0$, and defining,
\begin{equation}
I\equiv \int\limits_\mathrm{staple}d\mathbf{r}^\prime |\mathbf{r}^\prime|^2/{\cal A}, 
\label{eq:I}
\end{equation}
this gives,
\begin{equation}
\mathrm{compression:}\quad\left\{
\begin{array}{rl}
\mathbf{F}_i^\mathrm{dis}&=-k_d{\cal A}(\dot{\mathbf{r}}_i+\epsilon\mathbf{r}_i)\\[12pt]
\tau_i^\mathrm{dis}&=-k_d{\cal A}I\dot\theta_i.
\end{array}
\right.
\label{eq:compress}
\end{equation}

To model a system uniformly sheared in the $\mathbf{\hat x}$ direction at fixed strain rate $\dot\gamma$, we take as the average background velocity a uniform shear flow
\begin{equation}
\mathbf{v}_\mathrm{av}(\mathbf{r})=y\dot\gamma\mathbf{\hat x}
\end{equation}
and use Lees-Edwards boundary conditions \cite{Lees.JPhysC1972} on a box of fixed length $L$.  This gives,
\begin{equation}
\mathrm{shear:}\quad\left\{
\begin{array}{rl}
\mathbf{F}_i^\mathrm{dis}&=-k_d{\cal A}(\dot{\mathbf{r}}_i-y_i\dot\gamma\mathbf{\hat x}) \\[12pt]
\tau_i^\mathrm{dis}&=-k_d{\cal A}I[\dot\theta_i+\dot\gamma f(\theta_i)]
\end{array}
\right.
\label{eq:shear}
\end{equation}
where $\theta_i$ is the angle that the staple's spine makes with respect to the $\mathbf{\hat x}$ axis when the barbs are pointing downwards (so that $\theta_i=0$ in Fig.~\ref{fig:staple}), and 
\begin{equation}
f(\theta)\equiv \dfrac{1}{{\cal A}I}\int\limits_\mathrm{staple}d\mathbf{r}^\prime\> (y^\prime)^2.
\label{eq:ftheta}  
\end{equation}
Since the function $f(\theta)$ is always non-zero, an isolated particle (for which $\tau_i^\mathrm{el}=0$) will always undergo rotational motion in a shear flow, no matter what the particle shape.  
Except for particles with a particularly symmetric shape, this $f(\theta)$ will in general depend on the orientation  of the particle, and hence this rotational tumbling will be non-uniform.  Isolated particles will rotate most slowly at the orientations where $f(\theta)$ is minimum, and hence on average show a tendency to align at such orientations.  One goal of this work will be to investigate how inter-particle interactions may modify this rotational and orientational behavior of isolated particles.
Evaluating the integrals in Eqs.~(\ref{eq:I}) and (\ref{eq:ftheta}) for our staple shaped particle geometry, we find (see Appendix) 
\begin{equation}
f(\theta) = \frac{1 - C \cos 2 \theta}{2},
\label{eq:isolated_rotation}
\end{equation}
where $C$ is a function only of the barb to spine length ratio $\ell/w$. $C$ approaches 1 and -1 in the limiting cases where $\ell/w$ approaches 0 and $\infty$ respectively.  In both  limiting cases the staples approach a rod shape which strongly tends to align with the flow.

For both compression and shear, we will consider the case where the staple mass is sufficiently small that motion is in the overdamped limit, given by
\begin{equation}
\mathbf{F}_i^\mathrm{el}+\mathbf{F}_i^\mathrm{dis}=0,\qquad
\tau_i^\mathrm{el}+\tau_i^\mathrm{dis}=0.
\label{eq:overdamped}
\end{equation}
Substituting Eqs.~(\ref{eq:Fel}-\ref{eq:tauel}) and either Eq.~(\ref{eq:compress}) or  (\ref{eq:shear}) in the above then gives equations of motion for $\dot{\mathbf{r}}_i$ and $\dot\theta_i$, which can be numerically integrated to determine the translational and rotational motion of each of the staples.  
Henceforth we measure length in units of $D$, energy in units of $k_e$, and time in units of $t_0\equiv D^2k_d{\cal A}/k_e$.  
For our numerical integration we use Heun's modified Euler method with an integration step of $\Delta t=0.02 t_0$.

We will be interested in computing the pressure in our system as we vary the packing fraction and compression or shear rate.  We consider here only the elastic part of the pressure since it dominates over the kinetic and dissipative parts at the low strain rates we consider.  The elastic part of the stress tensor is given by \cite{BallGrinev},
\begin{equation}
{\bf P}=-\frac{1}{V}\sum_i\sum_{\mathrm{contacts}\>ab}\mathbf{r}_{iab}\otimes\mathbf{F}^{\mathrm{el}}_{ab}.
\end{equation}
The first sum is over all particles $i$, while the second sum is over all contacts $ab$ on particle $i$.  The moment arm $\mathbf{r}_{iab}$ is as defined following Eq.~(\ref{eq:tauel}), the contact force $\mathbf{F}^\mathrm{el}_{ab}$ is as defined following Eq.~(\ref{eq:Uel}), and $V$ is the total system volume.  The elastic part of the pressure for our two-dimensional system is then,
\begin{equation}
p = \frac{1}{2}\mathrm{Trace}[\mathbf{P}].
\end{equation}

\section{Results: Compression}

\begin{figure}
\includegraphics[width=0.9\columnwidth]{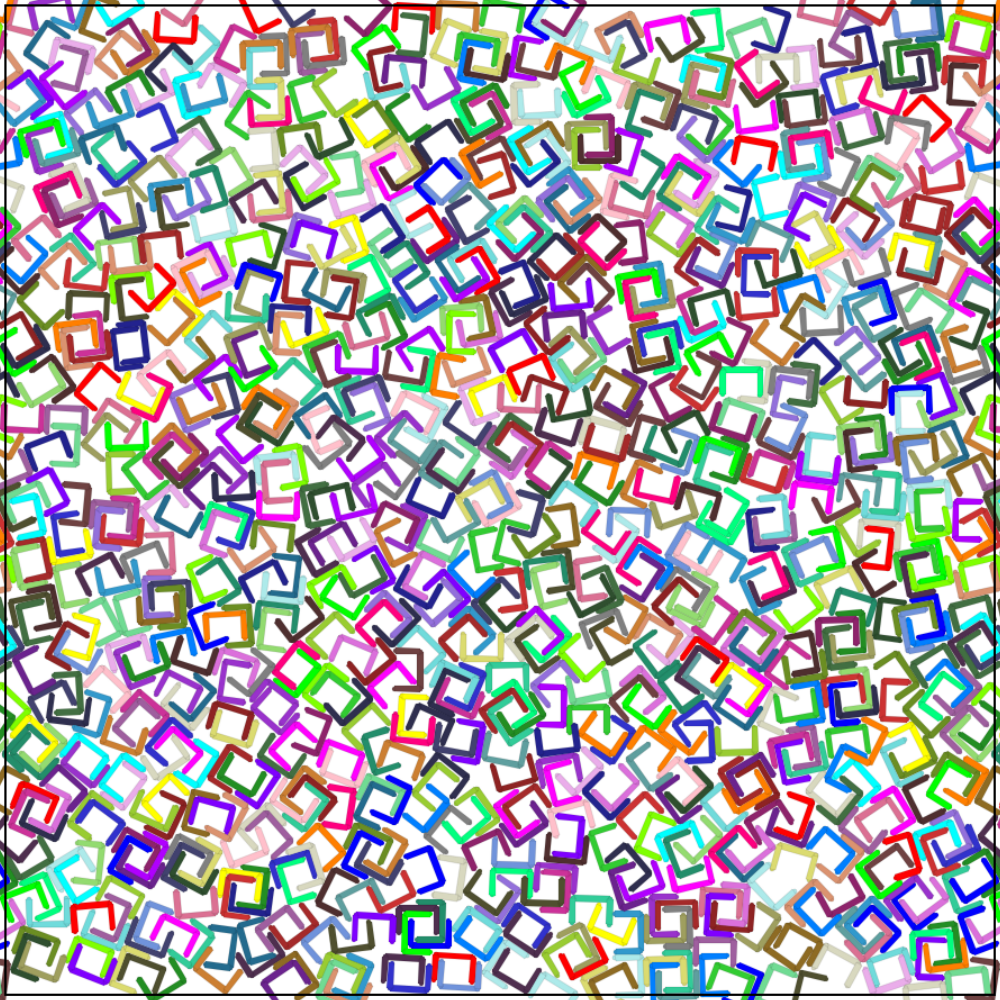}
\caption{\label{fig:snapshot}  A typical dense configuration of 1024 staples.  Here $\phi = 0.59$, and the system is being sheared at a rate $\dot{\gamma}=10^{-5}$.  Color is used merely to help distinguish different staples.}
\end{figure}

We first consider the jamming of the staples under uniform compression of the system. 
We begin our compression simulations starting from random dilute systems at packing fraction $\phi=0.2$.  
Initial states are chosen to have zero energy by placing staples one-by-one at random positions and orientations, rejecting placements which result in any staple overlaps.  
This allows us to prepare random systems without physically unrealistic effects such as staple axes penetrating through each other (see Fig.~\ref{fig:snapshot}). 
Examples of our compression simulations are shown in two animations (\href{http://www.pas.rochester.edu/~stte/staples/compress1.mpg}{Online Resource 1} and \href{http://www.pas.rochester.edu/~stte/staples/compress2.mpg}{Online Resource 2}), where we show compression from $\phi=0.20$ to $\phi=0.55$ for compression rates $\epsilon=10^{-5}$ and $10^{-7}$, respectively.
Numerical results presented below are averaged over at least 10 independent runs starting from different zero-energy configurations.

\begin{figure}
\centering
\includegraphics[width=0.9\columnwidth]{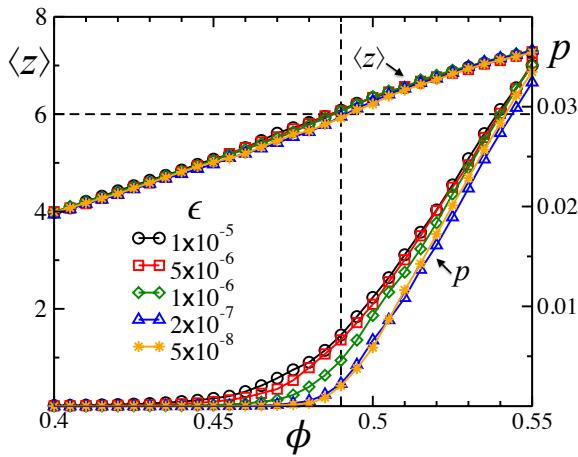}
\caption{\label{fig:compress}  Plot of pressure $p$ and average contact number $\langle z\rangle$ vs $\phi$ for several different compression rates $\epsilon$.  Jamming occurs at $\phi_J\approx 0.49$ when the isostatic condition $\langle z\rangle =6$ is satisfied. Pressure $p$ increases linearly above $\phi_J$.
}
\end{figure}

Figure~\ref{fig:compress} shows the elastic part of the pressure $p$ vs packing fraction $\phi$, for several different compression rates $\epsilon=1\times 10^{-5}$ to $5\times 10^{-8}$.  As $\phi$ increases, $p$ increases from zero.  As the compression rate $\epsilon$ decreases, the low-$\phi$ tail of $p$ sharpens up to give a jamming transition at $\phi_J\approx 0.49$.  As $\phi$ increases above $\phi_J$, pressure $p$ increases roughly linearly as has been found previously for frictionless disks and spheres with a harmonic interaction \cite{OHern.PRE2003}.

Jamming is often associated with the condition of isostaticity,  when the total number of degrees of freedom exactly equals the total number of constraints from the contact forces \cite{Alexander.PhysRep1998}. 
For frictionless particles, where contact forces are always normal to the surface at the point of contact, the isostatic condition is given by $Nd_f=Nz_\mathrm{iso}/2$, so $z_\mathrm{iso}=2d_f$.  
Here $d_f$ is the number of degrees of freedom per particle and one notes that each contact is shared by two particles.  
For spherically symmetric particles, which are invariant under rotation, only center of mass motion is relevant, so $d_f=d$ and $z_\mathrm{iso}=2d$.  
While frictionless disks and spheres have been clearly demonstrated to be isostatic at the jamming $\phi_J$ \cite{OHern.PRE2003}, ellipsoidal particles have been found to be hypostatic at jamming, with $\langle z\rangle <2d_f$ \cite{Donev.Science2004,Donev.PRE2007,Zeravcic.EPL2009,Mailman.PRL2009}.  
It has further been argued that smooth convex shaped particles will in general be hypostatic at jamming \cite{Donev.PRE2007,Roux.PRE2000}. 
However our staples are concave, and so it remains in question whether isostaticity describes the state of staples at jamming. 

For particles with no rotational symmetries, such as our staples, there are $d_f= d(d+1)/2$ total translational and rotational degrees of freedom per particle \cite{Roux.PRE2000}.  
Thus the isostatic condition for our staples in $d=2$ dimensions is $z_\mathrm{iso}=2d_f=6$.  
Note that, since our particles are concave, the same two neighboring staples may contact each other at more than one point, and in fact may share up to 4 different contacts.  
Therefore the average number of contacts per staple $\langle z\rangle$, is in general greater than the average number of neighbors each staple is in contact with, as has been observed for other non-convex particles \cite{Papanikolaou.PRL2013}.

In Fig.~\ref{fig:compress} we show $\langle z\rangle$ vs $\phi$ as we compress with different rates $\epsilon$.  Comparing the curves of $\langle z\rangle$ against the curves of pressure $p$, we see that isostaticity $\langle z\rangle=6$ does indeed seem to hold at the jamming transition $\phi_J\approx 0.49$. By fitting the linear portion of the pressure curve at our smallest compression rate $\epsilon=5\times 10^{-8}$, and extrapolating to zero, we find that $\phi_J$ where this pressure vanishes agrees with the isostatic packing fraction where $\langle z\rangle=6$.

We have also found that the jamming packing fraction $\phi_J$ depends slightly on the initial packing fraction at which the starting zero-energy configurations are prepared.  
Systems which were initialized at very dilute packing fractions, $\phi\le0.2$, all jammed at the same $\phi_J\approx 0.49$.  However, when the packing fraction of the initial state increased, the subsequent jamming $\phi_J$ also slightly increased.  
For configurations prepared at an initial $\phi=0.3$, a $\phi_J\approx 0.5$ was observed. 
A similar dependence of $\phi_J$ on the ensemble of initial states from which compression begins was found for frictionless spheres and disks \cite{Chaudhuri.PRL2010,Vagberg.PRE2011}.

\section{Results: Shear}

\begin{figure}
\centering
\includegraphics[width=0.9\columnwidth]{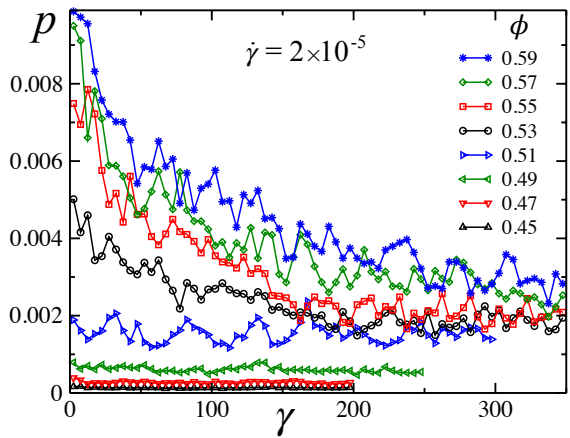}
\caption{\label{fig:poft} Plot of pressure $p$ as a function of the net shear strain $\gamma=\dot\gamma t$ for several different packing fractions $\phi$, at a strain rate of $\dot\gamma=2\times 10^{-5}$.  Each point represents a local average over a strain window of $\Delta\gamma=5$.}
\end{figure}

We now consider the jamming of the staples under the application of a uniform applied shear strain rate $\dot\gamma$.  
We investigate systems with packing fractions in the interval $\phi\in[0.45,0.59]$, and shear at fixed $\phi$ for a range of strain rates $\dot\gamma=2\times 10^{-4}$ to $2\times 10^{-6}$.
For each value of $\phi$ and $\dot{\gamma}$ we initialize the system by starting with a different zero-energy configuration at a dilute packing fraction, and then compressing to the desired packing fraction $\phi$ before shearing.   
Examples of our shearing simulations are shown in two animations (\href{http://www.pas.rochester.edu/~stte/staples/shear1.mpg}{Online Resource 3} and \href{http://www.pas.rochester.edu/~stte/staples/shear2.mpg}{Online Resource 4}), where we show shearing at two different packing fractions $\phi=0.51$ and $\phi=0.59$, both with a strain rate $\dot\gamma=5\times 10^{-6}$.
While we expect that the system, when sheared long enough, will eventually lose memory of its initial configuration \cite{Vagberg.PRE2011}, we find that memory of the initial configuration, particularly at denser $\phi$ and slower $\dot\gamma$, can persist for quite long strains.  
In Fig.~\ref{fig:poft} we plot the pressure $p$ as a function of the net shear strain $\gamma=\dot\gamma t$, for several different packing fractions $\phi$, at a strain rate of $\dot\gamma=2\times 10^{-5}$.  Each point $p(\gamma)$ represents a local average over a strain window of $\Delta\gamma=5$ centered about the strain $\gamma$.
We see clearly that the relaxation time increases as $\phi$ increases, and at the highest $\phi$ it takes a total strain of $\gamma > 150$ to reach the steady state.  
Our results below are obtained by waiting until the system has reached steady state, and then averaging over a further total strain $\gamma \geq 100$.  For $\phi\geq 0.51$, where fluctuations are large, we further average over two independent runs.

\subsection{Pressure and Jamming}

\begin{figure}
\centering
\subfigure{
\includegraphics[width=0.9\columnwidth]{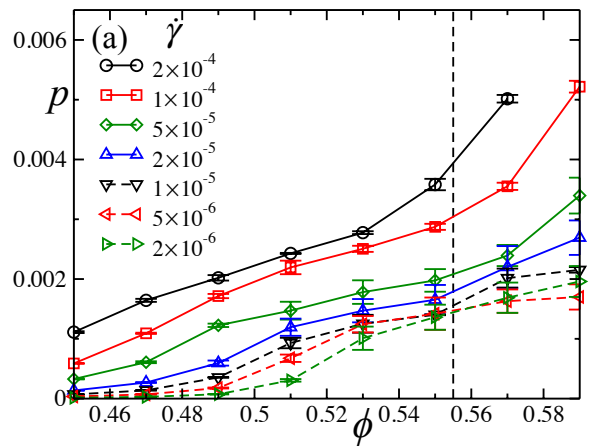}}
\subfigure{
\includegraphics[width=0.9\columnwidth]{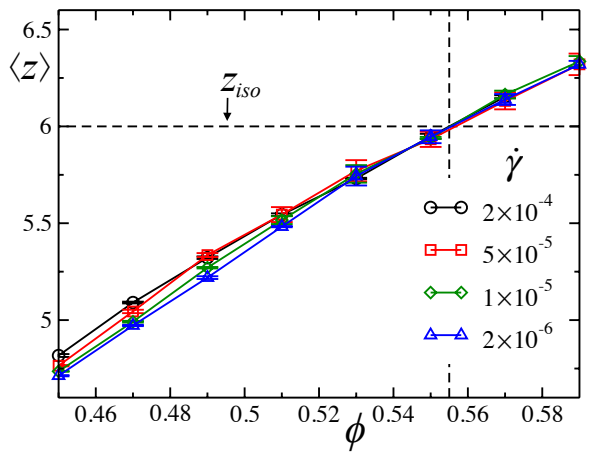}}
\caption{\label{fig:shear_pz}  Plots of (a) pressure $p$ and (b) average contact
number $\langle z\rangle$ versus $\phi$ in steady-state shear at several
different values of uniform shear strain rate $\dot{\gamma}$. Dashed vertical lines indicate the value of $\phi$ at which the isostatic condition, $\langle z\rangle=6$, occurs.
}
\end{figure}

In Fig.~\ref{fig:shear_pz}a we show the resulting steady-state average of the pressure $p$ as a function of the packing fraction $\phi$. 
As with compression, we find that systems sheared at slower rates  show pressure curves that shift towards higher packing fractions $\phi$.  In the limit $\dot\gamma\to 0$, the measured $p$ represents the pressure along the yield stress curve;
we expect in principle to see $p$ vanish for all $\phi<\phi_J$, and then rise to finite values above $\phi_J$.
It appears that the pressure begins to converge at finite values when $\phi \geq 0.53$.  However it is difficult to estimate the precise value of the shear-driven jamming $\phi_J$.  Our data are not accurate enough, nor our system large enough, to do a critical scaling analysis to determine $\phi_J$, as has been done for the case of frictionless disks \cite{Olsson.PRE2011}.

If we believe that, as in compression, shear-driven jamming will occur when the system satisfies the isostatic condition, then we see from the plot of $\langle z\rangle$ vs $\phi$ in Fig.~\ref{fig:shear_pz}b that this occurs when $\phi \approx 0.555$.  Comparing with Fig.~\ref{fig:shear_pz}a (note vertical dashed lines) we see that this occurs noticeably  above the $\phi$ at which the pressure increases above its small $\phi$ tail, somewhere around $\phi = 0.52$.  

\begin{figure}
\centering
\includegraphics[width=0.95\columnwidth]{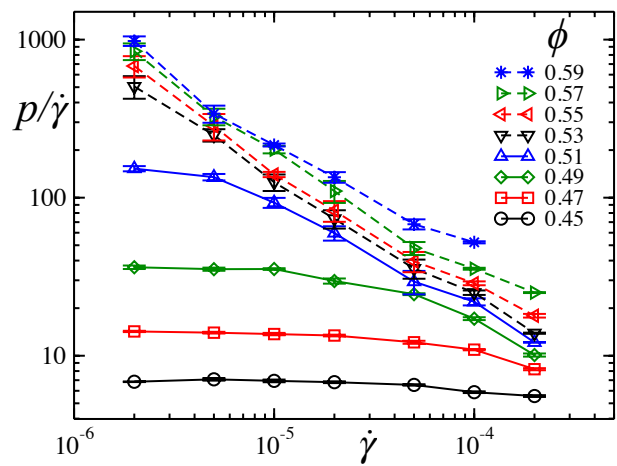}
\caption{\label{fig:shear_viscosity}  Plot of $p/\dot{\gamma}$ vs $\dot{\gamma}$ for several different fixed values of $\phi$.  When the curves saturate to a constant value as $\dot\gamma\to 0$, it indicates that the pressure $p\propto \dot\gamma$, and so $p$ tends to zero as $\dot\gamma\to 0$, and hence the system is unjammed.}
\end{figure}

As another way to look for the limiting $\dot\gamma\to 0$ behavior, and so determine $\phi_J$, we plot in Fig.~\ref{fig:shear_viscosity} the pressure analog of viscosity $p/\dot{\gamma}$ vs $\dot\gamma$ at various fixed values of $\phi$.  We expect that $p/\dot\gamma$ will saturate to a finite value as $\dot\gamma\to 0$ for all $\phi<\phi_J$.  But since above $\phi_J$ the system supports a finite stress even as the shear rate approaches zero, we expect $p/\dot\gamma$ must
diverge as $\dot\gamma\to 0$ for $\phi>\phi_J$.
At low $\phi\le 0.51$ in Fig.~\ref{fig:shear_viscosity} we clearly see the expected plateau to a finite value as $\dot\gamma$ decreases.  
As $\dot\gamma$ increases for these low $\phi$, we see the shear thinning behavior (decreasing $p/\dot\gamma$) that is typical of overdamped, soft, frictionless granular materials \cite{Olsson.PRE2011,Roux.RheoActa2009}.  At higher $\phi$ we see a continuing increase in $p/\dot\gamma$ as $\dot\gamma$ decreases.  
Our data suggests that the crossover between these two different limiting behaviors occurs at roughly $\phi\approx 0.52$.  
However one cannot say with confidence whether the curves at $\phi>0.52$ will continue to increase, or may bend over to saturate to a finite value, as $\dot\gamma$ decreases to smaller values than we have been able to simulate.  
Our value $\phi\approx 0.52$ should therefore be taken as a lower bound for the shear-driven $\phi_J$.
Though we can only give a lower boud for the shear-driven $\phi_J$, we note that it is clearly larger  than the compression-driven $\phi_J$, as has also been found for frictionless disks \cite{Vagberg.PRE2011,Vagberg.PRE2011b}.

\begin{figure}
\centering
\includegraphics[width=0.95\columnwidth]{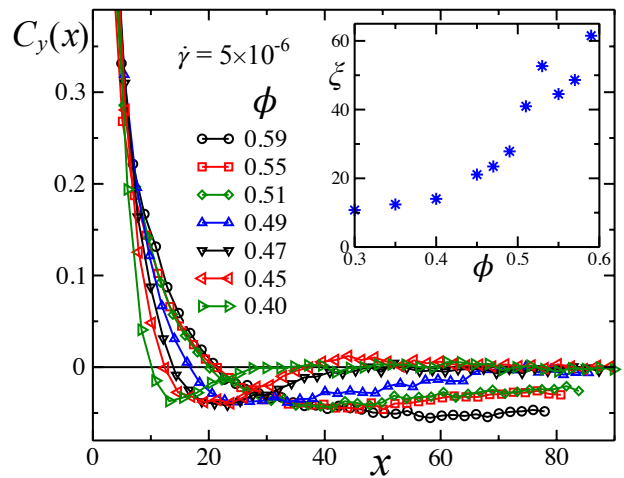}
\caption{\label{fig:Cy}Transverse velocity correlation, $C_y(x)\equiv\langle v_y(0)v_y(x)\rangle,$ vs position $x$, for various packing fractions $\phi$ at fixed shear strain rate $\dot\gamma=5\times 10^{-6}$. The location of the minimum of $C_y(x)$ determines the translational correlation length $\xi$, which is plotted vs $\phi$ in the inset. For computing the correlation, we have used a bin width of $\Delta x =2D$, equal to twice the staple diameter, for measuring staple separations.
}
\end{figure}

For another attempt to locate the shear driven jamming transition, we consider the transverse velocity correlation function,
\begin{equation}
C_y(x)\equiv\langle v_y(0)v_y(x)\rangle,
\end{equation}
where $v_{y}(\mathbf{r})$ is the $y-$component of the center of mass velocity, transverse to the direction of flow,  of a particle at position $\mathbf{r}$.  For circular frictionless disks, $C_y(x)$ is known to have a well defined minimum at a distance $x=\xi$, where $\xi$ has been interpreted as a translational correlation length; $\xi$ increases and diverges as one approaches the jamming transition, $\phi\to\phi_J$ and $\dot\gamma\to 0$ \cite{Olsson.PRL2007}.  In Fig.~\ref{fig:Cy} we plot $C_y(x)$ vs $x$ for our staples, for $x\in[0, L/2]$, showing data for several different packing fractions $\phi$ at a fixed slow shear strain rate $\dot\gamma=5\times 10^{-6}$.  We find the same qualitative  behavior as for disks: 
the length scale $\xi$ locating the minimum of $C_y(x)$ increases steadily as $\phi$ increases.  
We plot this $\xi$ vs $\phi$ as the inset to Fig.~\ref{fig:Cy}.
We note that the longest tip-to-tip distance along the staple's spine is $A+3D=7$, so at our lowest packing fraction $\xi\approx 10$ is about one and a half staple lengths.  At $\phi=0.55$, $\xi\approx 45$, or about 6 staple lengths.  And at our largest $\phi=0.59$, $\xi\approx 60$, but the minimum is very shallow.  We note that the finite system size $L$, and the finite strain rate $\dot\gamma$, both act to reduce the correlation length from its value in the infinite size, vanishing strain rate, limit.  Thus while we have clear evidence for a growing, macroscopically large, translational correlation length as $\phi$ increases, it is difficult to infer from this data a clear value for $\phi_J$.  Being unable to give a precise value for $\phi_J$, our results therefore remain inconclusive as to whether the system at the shear driven jamming point $\phi_J$ is isostatic or slightly hypostatic.

\subsection{Orientational Ordering } 

\begin{figure}
\centering
\includegraphics[width=0.95\columnwidth]{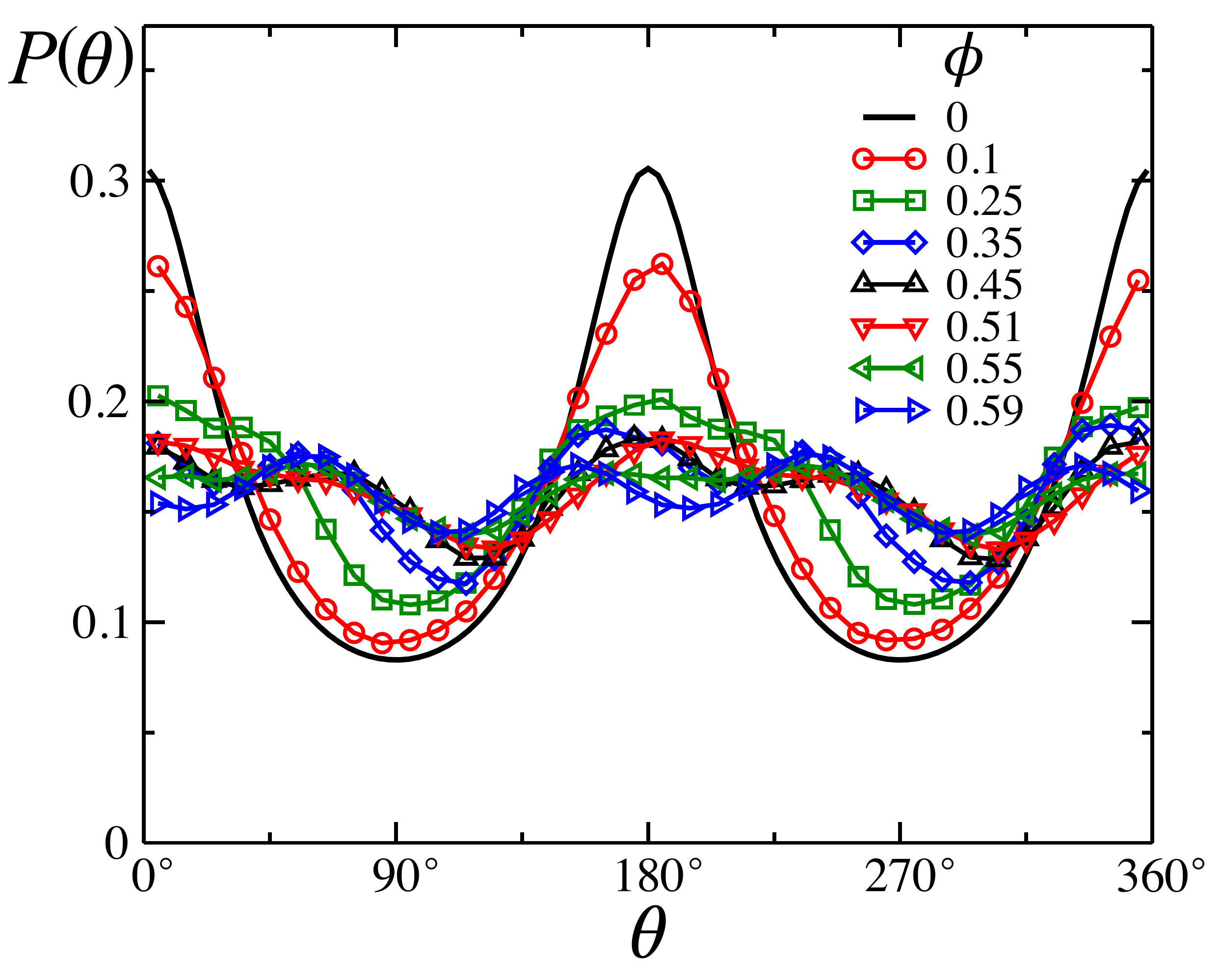}
\caption{\label{fig:pdf_theta}  The probability density functions for both an isolated staple in a uniform shear flow (solid black line) and for sheared systems of staples at packing fractions $\phi \in [0.1,0.59]$.  
The shear rate is $\dot{\gamma} = 5\times10^{-5}$ for $\phi < 0.45$ and $\dot{\gamma} = 10^{-5}$ for $\phi \geq 0.45$. At low $\phi$ the staples prefer to orient with their spines parallel to the flow direction.}
\end{figure}

A new physical effect associated with non-spherically symmetric particles in a shear flow, is that the flow will cause the particles to tumble.
The rotational motion of a staple is governed by the torque balance of Eq.~(\ref{eq:overdamped}).  For an {\em isolated} staple $\tau_i^\mathrm{el}=0$, and the rotation is then determined by the condition $\tau_i^\mathrm{dis}=0$, or from Eq.~(\ref{eq:shear}), $\dot{\theta_i}=-\dot{\gamma}f(\theta_i)$, with $f(\theta)$ given by Eq.~(\ref{eq:isolated_rotation}).  Since $f(\theta)>0$, the rotation is clockwise for positive $\dot\gamma$.  Since $f(\theta)$ varies with $\theta$, the rotation is non-uniform, being slowest when $f(\theta)$ has its minimum at $\theta=0$ or $\pi$.  An isolated staple thus spends more time with its spine oriented parallel to the direction of the shear flow.  One can see this more explicitly by directly computing the probability density for an isolated staple to be oriented at angle $\theta_0$,
\begin{equation}
\begin{aligned}
P(\theta_0)&=\frac{\bar\omega}{2\pi}\int_0^{2\pi/\bar\omega}  \delta(\theta(t)-\theta_0) dt\\[12pt]
&=\frac{\bar\omega}{2\pi}\int_0^{2\pi}  \frac{\delta(\theta-\theta_0)}{|\dot\theta|}d\theta=\dfrac{\bar\omega}{2\pi\dot\gamma f(\theta_0)},
\end{aligned}
\end{equation}
where $\delta(\theta)$ is the Dirac delta function, $2\pi/\bar\omega$ is the period of one rotation, with $\bar\omega$ the magnitude of the average angular velocity; the integral just gives the fraction of one period that the particle spends at any particular angle $\theta_0$.  Normalization of $P(\theta)$ then determines,
\begin{equation}
\frac{2\pi}{\bar\omega}=\int_0^{2\pi}\frac{d\theta}{\dot\gamma f(\theta)}=\frac{4\pi}{\dot\gamma\sqrt{1-C^2}}.
\end{equation}
The average angular velocity of an {\em isolated} staple is thus 
\begin{equation}
\frac{\bar\omega}{\dot\gamma}=\frac{1}{2}\sqrt{1-C^2}\qquad\mathrm{(for\>an\>isolated\>staple)},
\label{eq:wisolated}
\end{equation}
varying from 0 to a maximum value of $1/2$ (corresponding to uniform rotation) depending on the barb to spine ratio $\ell/w$ (see Appendix).  The distribution $P(\theta)$ for an isolated staple with $\ell/w=2/3$ is shown as the solid black line in Fig.~\ref{fig:pdf_theta}.

When particles are packed together and allowed to interact, the resulting collisions will give rise to a non-zero $\tau_i^\mathrm{el}$.  One may naively expect that collisions will effect the rotation of a particular particle in two possible ways: (i) the excluded volume occupied by other particles may block rotation, leading to a lower average angular velocity $\bar\omega$ and a possible increase in particle alignment, or (ii) the collisions may act like random kicks, knocking the particle out of its preferred orientations at $\theta_i=0,\pi$, thus increasing $\bar\omega$ and reducing particle alignment.  We will find that for staples, the second effect appears to dominate.


In Fig.~\ref{fig:pdf_theta} we show our numerically computed $P(\theta)$ for several different values of the packing fraction $\phi$.  We find there to be little to no dependence on the shear rate $\dot\gamma$ for the rates considered here, so we show results for only one specific slow $\dot\gamma$, depending on $\phi$.
We see that as $\phi$ increases the collisions decrease the likelihood of a particle to be oriented at $\theta=0$ or $\pi$, and generally act to flatten the distribution compared to that of an isolated staple.  
This is in contrast to what has been observed for sheared rods \cite{Campbell,Guo,Borzsonyi.PRL2012,Borzsonyi2,Guo.PhyFluids2013}, where $P(\theta)$ sharpens as $\phi$ increases.  
As $\phi$ approaches $0.59$, we see the development of four preferred orientations, all shifted away from $\theta=0$ and $\pi$.

We can quantify the information contained in $P(\theta)$ by computing orientational order parameters.  We consider three possible cases: (i) vectorial ordering, where the staple aligns in a particular preferred direction $\theta_1$; (ii) nematic ordering, where the staple's spine alignes in a particular direction (the ``director") $\theta_2$, independent of the direction of the barbs; (iii) tetratic ordering, where either the staple's spine or barbs align in a particular direction (the ``bidirector") $\theta_4$. Tetratic ordering is suggested by the tendency of dense staples to nest within each other at orthogonal orientations, as seen in Fig.~\ref{fig:snapshot}, and the appearance of the four preferred orientations in $P(\theta)$ as $\phi$ increases, as seen in Fig.~\ref{fig:pdf_theta}.  We denote these three cases by $m=1,2,4$, respectively.  The ordering direction is then $\theta_m$, and the corresponding scalar order parameter is $S_m$.

\begin{figure}[h!]
\centering
\includegraphics[width=0.9\columnwidth]{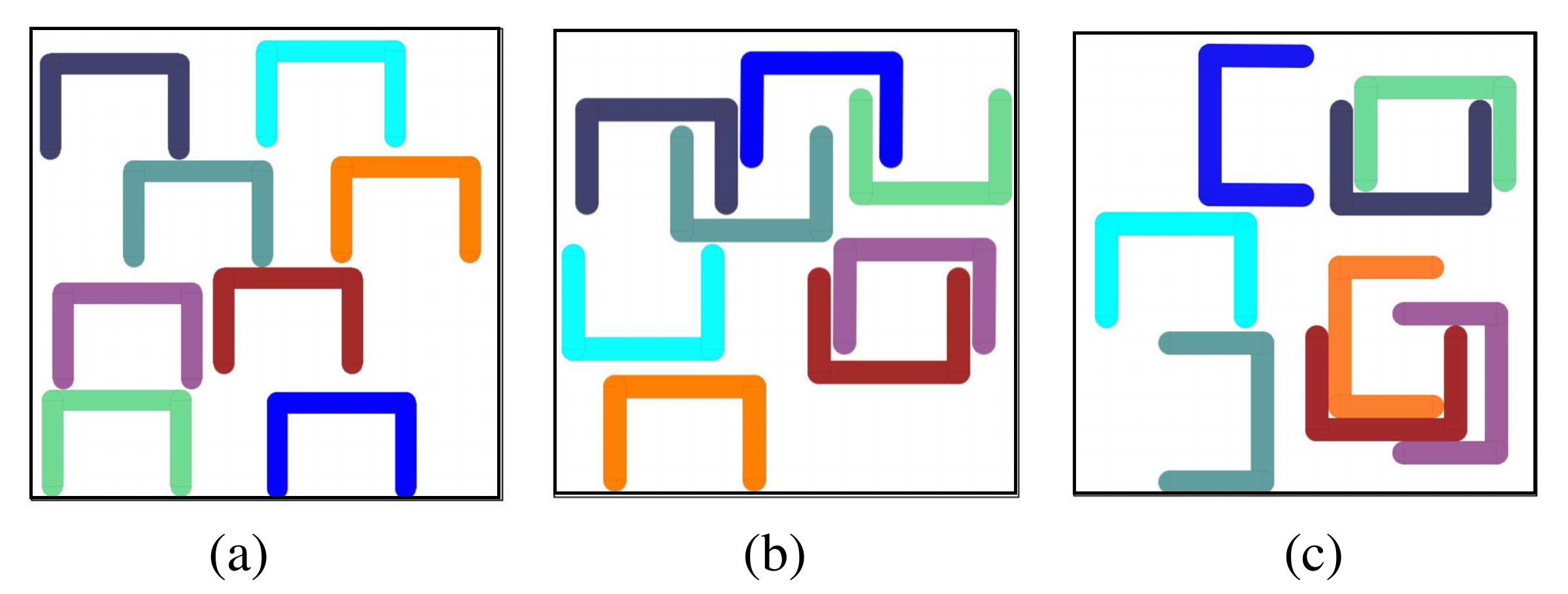}
\caption{\label{fig:samples}  
Idealized examples of perfect orientational ordering: (a) vectorial ordering with $S_1=1$ at orientation $\theta_1=0$, (b) nematic ordering with $S_1=0$ but $S_2=1$, with director oriented at $\theta_2=0$, (c) tetratic ordering with $S_1=S_2=0$ but $S_4=1$, with bidirector oriented at $\theta_4=0$.
}
\end{figure}

To illustrate these three different forms of orientational order, we show in Fig.~\ref{fig:samples} examples of systems with perfect vectorial, nematic, and tetratic ordering; these are only idealized sketches, {\em not} actual configurations encountered in our simulations.

For a two-dimensional system, one may easily compute both $S_m$ and $\theta_m$ by \cite{Donev.PRB2005},
\begin{equation}
S_m=\max_{\theta_m}\left[\,\langle\cos(m(\theta_i-\theta_m)\rangle\,\right],
\end{equation}
where the $\theta_m$ that maximizes the average is the ordering direction.  One then finds,
\begin{equation}
\tan(m\theta_m)=\frac{\langle\sin(m\theta_i)\rangle}{\langle\cos(m\theta_i)\rangle},
\end{equation}
and
\begin{equation}
S_m=\sqrt{\langle\cos(m\theta_i)\rangle^2+\langle\sin(m\theta_i)\rangle^2}.
\end{equation}

\begin{figure}
\centering
\subfigure{
\includegraphics[width=0.9\columnwidth]{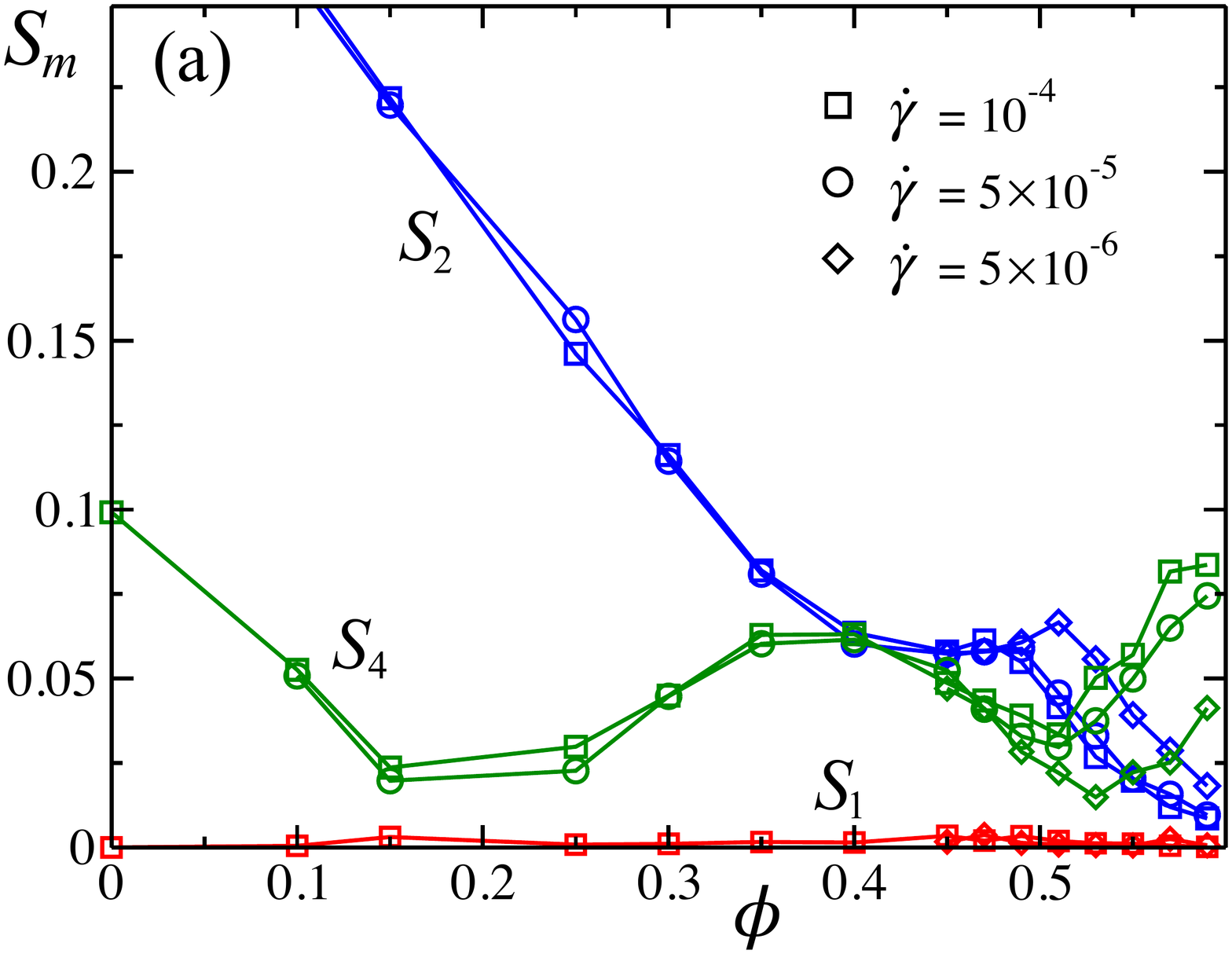}}
\subfigure{
\includegraphics[width=0.9\columnwidth]{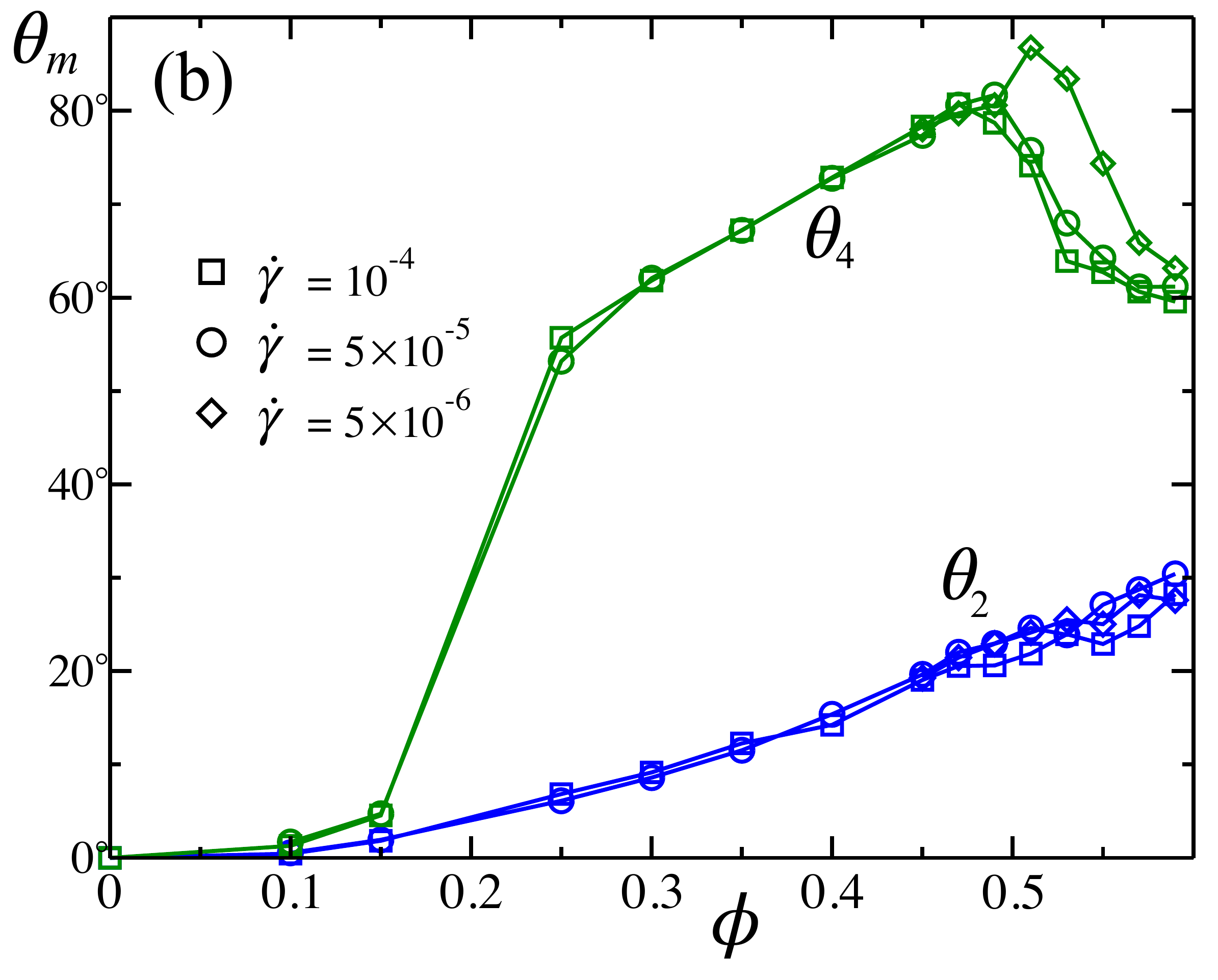}}
\caption{\label{fig:order}  Plots of (a) orientational order parameters $S_m$ for vectorial ($m=1$), nematic ($m=2$) and tetratic ($m=4$) ordering, and (b) ordering angle $\theta_m$, for nematic and tetratic ordering.  Results are shown for the three different shear strain rates $\dot\gamma=10^{-4}$ (squares), $5\times 10^{-5}$ (circles), and $5\times 10^{-6}$ (diamonds).
}
\end{figure}

In Fig.~\ref{fig:order}a we plot the order parameters $S_m$ vs $\phi$, for $m=1,2,4$, for three different shear strain rates $\dot\gamma$.
In Fig.~\ref{fig:order}b we plot the corresponding ordering angles $\theta_m$.  We see the following behavior.  There is no vectorial ordering, with $S_1\approx 0$ for all $\phi$; hence there is no meaningful $\theta_1$.  The nematic ordering $S_2$ decreases steadily as $\phi$ increases, saturating to a plateau just below jamming, and then decreasing to zero as the system goes above jamming.  The corresponding ordering angle of the director $\theta_2$ steadily increases from zero (aligned with flow) to $\theta_2\approx 30^\circ$ near jamming.  A similar orienting at positive $\theta$ with respect to the flow direction has previously been observed for rod shaped particles \cite{Campbell,Guo,Borzsonyi.PRL2012,Borzsonyi2,Guo.PhyFluids2013}.  The tetratic ordering $S_4$ behaves non-monotonically at low $\phi$, then decreases towards zero as jamming is approached from below, only to increase as $\phi$ increases above jamming.  At our highest $\phi=0.59$, $\theta_4\approx 60^\circ$, which modulus the $90^\circ$ periodicity of tetratic ordering, agrees with the location of the four peaks in $P(\theta)$ seen in Fig.~\ref{fig:pdf_theta} at this $\phi$.  The jamming transition thus appears to be accompanied by a vanishing of nematic ordering and the increase in tetratic ordering.  As might be expected, sensitivity to the strain rate $\dot\gamma$ sets in near the jamming transition, showing a slight shift of the transition from nematic to tetratic ordering to higher $\phi$ as $\dot\gamma$ decreases.

To see whether or not the orientational ordering is a consequence of interaction induced collective behavior,
we now consider the orientational ordering correlation functions,
\begin{equation}
G_m(\mathbf{r})\equiv\langle\cos (m(\theta(0)-\theta(\mathbf{r}))\rangle,
\end{equation}
where $\theta(\mathbf{r})$ denotes the orientation angle of a particle at position $\mathbf{r}$.

In general $G_m(\mathbf{r})$ will approach a finite constant $G_m(\infty)$ as $|\mathbf{r}|\to\infty$.  We therefore define a modified correlation that decays to zero,
\begin{equation}
\tilde G_m(\mathbf{r})\equiv G_m(\mathbf{r}) - G_m(\infty).
\end{equation}
When infinitely far apart, particles $i$ and $j$ are uncorrelated, and so one can write,
\begin{equation}
\begin{aligned}
G_m(\infty) &=\! \int_0^{2\pi} \!\!\!\! \int_0^{2\pi} P(\theta_i) P(\theta_j) \cos m(\theta_i - \theta_j) \mathrm{d}\theta_i \mathrm{d}\theta_j\\[12pt]
&=S_m^2
\end{aligned}
\end{equation}

\begin{figure}
\centering
\subfigure{
\includegraphics[width=0.9\columnwidth]{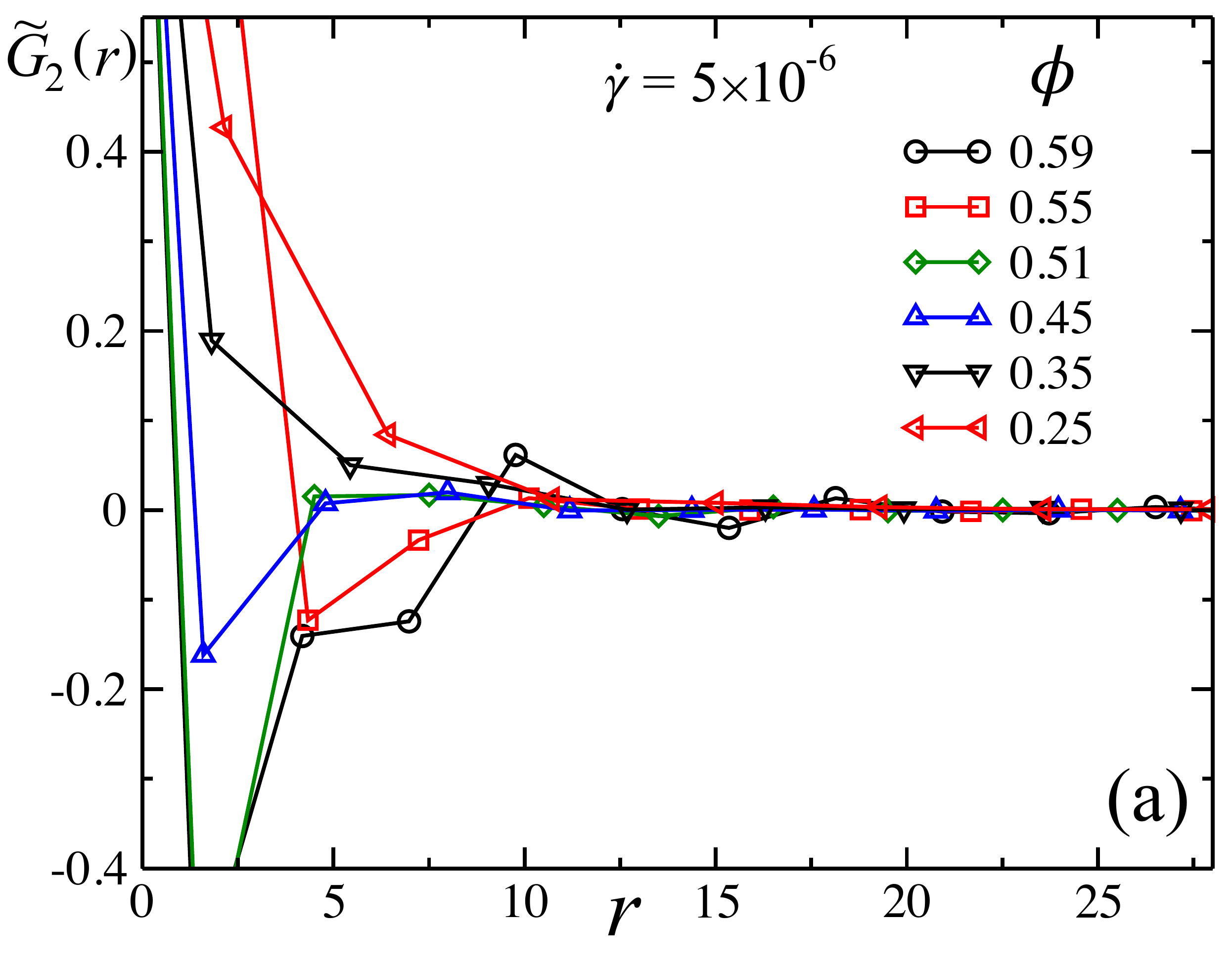}}
\subfigure{
\includegraphics[width=0.9\columnwidth]{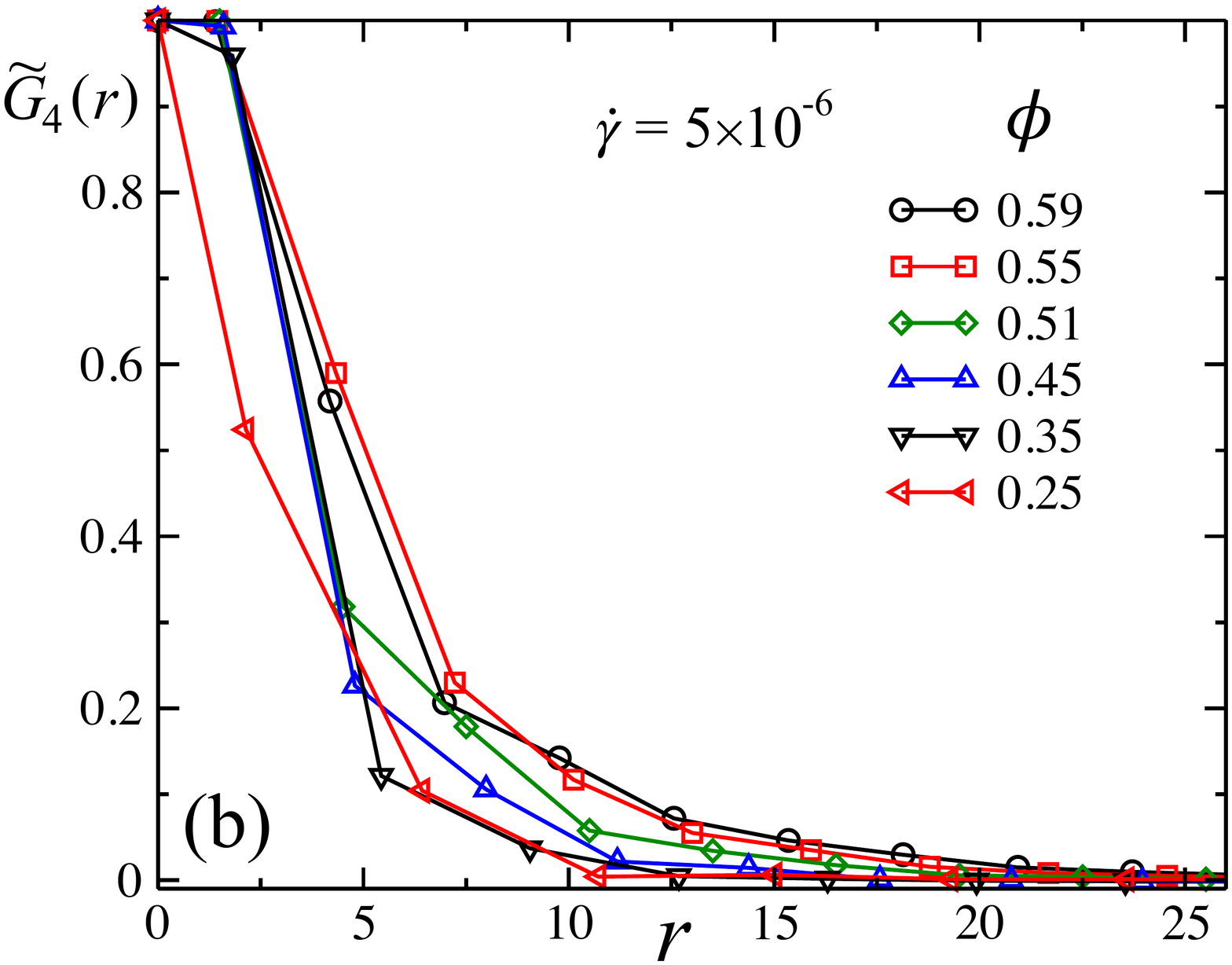}}
\caption{\label{fig:Gr}  Angular averaged correlations functions $\tilde G_m(r)$ vs radial distance $r$ for (a) nematic ordering, $m=2$ and (b) tetratic ordering, $m=4$, for several different packing fractions $\phi$ at a fixed shear strain rate $\dot\gamma=5\times 10^{-6}$.  For computing the correlation, we have used a bin width of $\Delta r =2D$, equal to twice the staple diameter, for measuring staple separations.
}
\end{figure}

Because we have observed no noticeable directional dependence to the correlation functions, we have computed their angular average $\tilde G_m(r)$, averaging over all separations of fixed magnitude $r=|\mathbf{r}|$.
In Fig.~\ref{fig:Gr} we plot $\tilde G_2(r)$ and $\tilde G_4(r)$ vs the radial distance $r$ for several different packing fractions $\phi$ at a fixed $\dot\gamma=5\times 10^{-6}$ (as with $S_m$ and $\theta_m$, we find little dependence of $\tilde G_m$ on $\dot\gamma$).  In both nematic and tetratic cases the corresponding orientational correlation length, defined as the length scale on which $\tilde G_m(r)$ approaches zero, never gets more than one staple's length $\sim 7$.

We thus conclude that the presence of a finite orientational ordering $S_m>0$  is a consequence of the shear strain rate $\dot\gamma$ serving as an ordering field, rather than any cooperative behavior among large numbers of particles.  As $\phi$ increases, the interactions with the other particles act like a disordering noise that reduces the effect of the strain ordering field and causes $S_m$ to decrease.  The growth in tetratic order above jamming would appear to be an interaction effect, but of only a local nature. 

\subsection{Rotation of Particles}

For more insight into the reason that  orientational ordering generally decreases as $\phi$ increases,  we look at the average angular velocity,  $\langle \dot{\theta}_i \rangle$.  Since the average rotation is clockwise for $\dot\gamma >0$, we have $\langle\dot\theta_i\rangle <0$ at all $\phi$. We therefore define $\bar\omega = |\langle\dot\theta_i\rangle|$ as the magnitude of this angular velocity.

\begin{figure}
\centering
\includegraphics[width=0.95\columnwidth]{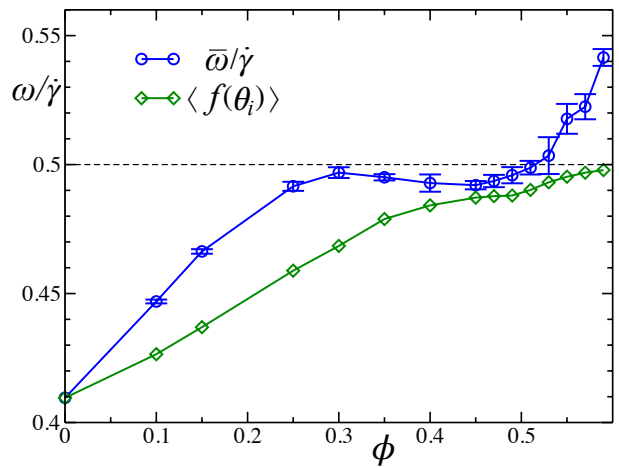}
\caption{\label{fig:angular_velocity}  The magnitude of the average angular rotation $\bar\omega/\dot{\gamma}$, and the contribution to this rotation from the background shear velocity $\langle f(\theta_i) \rangle$.  Both are independent of the shear rate $\dot{\gamma}$, and are averaged over independent runs with shear rates $\dot{\gamma} \in [5\times10^{-5},5\times10^{-6}]$.}
\end{figure}

At the low strain rates considered here, we find $\bar\omega$ strictly proportional to the strain rate $\dot\gamma$, both below and above jamming. Hence in Fig.~\ref{fig:angular_velocity} we plot the dimensionless $\bar\omega/\dot\gamma$ vs $\phi$, which is independent of $\dot\gamma$.  We average our results over strain rates in the interval $\dot\gamma\in[5\times 10^{-5}, 5\times 10^{-6}]$ to give greater statistical accuracy.  We see that as $\phi$ increases from zero, $\bar\omega/\dot\gamma$ increases from the value $\frac{1}{2}\sqrt{1-C^2}$ given by Eq.~(\ref{eq:wisolated}) for an isolated staple, to plateau at roughly $\bar\omega/\dot\gamma\approx 1/2$ at $\phi\approx 0.3$, well below jamming. As $\phi$ increases above the jamming $\phi_J\approx 0.52$, $\bar\omega/\dot\gamma$ increases above $1/2$.  This increase in $\bar\omega/\dot\gamma$ as $\phi$ increases is in stark contrast to the behavior of frictionless rod shaped particles, where we have found that $\bar\omega/\dot\gamma$ monotonically {\em decreases} as $\phi$ increases \cite{DVH}.

In Fig.~\ref{fig:angular_velocity} we also plot $\langle f(\theta_i)\rangle$. From Eq.~(\ref{eq:shear}) we see that $\langle f(\theta_i)\rangle$ is the contribution to $\bar\omega/\dot\gamma$ arising from the dissipative torque due to the background shear flow.  We see that $\langle f(\theta_i)\rangle$ rises from its value for the isolated staple at $\phi=0$ to the value $1/2$ as $\phi$ increases.  The value $1/2$ characterizes the situation where all angles $\theta_i$ are equally likely, and rotation is on average uniform.  The rise in $\langle f(\theta_i)\rangle$ to 1/2 thus reflects the flattening of the $P(\theta)$ distribution seen in Fig.~\ref{fig:pdf_theta}.  We see from Fig.~\ref{fig:angular_velocity} that $\langle f(\theta_i)\rangle$ gives the dominant contribution to $\bar\omega/\dot\gamma$.

The difference between the curves in Fig.~\ref{fig:angular_velocity},
\begin{equation}
\bar\omega/\dot\gamma-\langle f(\theta)\rangle = -\langle \tau_i^\mathrm{el}\rangle/(k_d{\cal A}I\dot\gamma), 
\end{equation}
is the contribution to the angular rotation from the elastic collisions, as follows from Eqs.~(\ref{eq:shear}) and (\ref{eq:overdamped}).
Thus, for staples, the elastic collisions always {\em increase} the average rate of rotation.  In particular, the rise in $\bar\omega/\dot\gamma$ above 1/2, as $\phi$ increases above jamming, is due to torquing from the elastic collisions.


We can get a more in depth look at the contacts between staples by examining the fabric tensor \cite{Zhang.GM2010}, which is defined as
\begin{equation}
\mathbf{R} = \frac{1}{N} \sum_{i} \sum_{\mathrm{contacts\,}ab} \mathbf{r}_{i ab} \otimes \mathbf{r}_{i ab},
\end{equation}
where the first sum is over all particles $i$, the second sum is over all contacts $ab$ that the component spherocylinders $a$ of $i$ make with the spherocylinders $b$ of other particles $j$, and $\mathbf{r}_{iab}$ is the displacement from the center of mass of staple $i$ to the point of contact $ab$.
If $\lambda_1 > \lambda_2$ are the two eigenvalues of $\mathbf{R}$, then the trace $\lambda_1+\lambda_2$ is simply the average contact number $\langle z \rangle$.  The difference $\Delta\lambda=\lambda_1-\lambda_2$ is a measure of the directional anisotropy of the contacts in the system. A large $\Delta\lambda$ means more contacts oriented along the direction of the eigenvector associated with $\lambda_1$, and fewer contacts in the orthogonal direction; vanishing $\Delta\lambda$ means contacts are distributed isotropically.

\begin{figure}
\centering
\includegraphics[width=0.95\columnwidth]{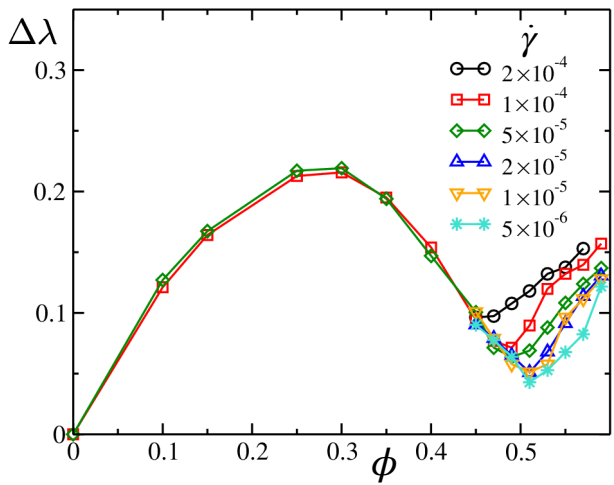}
\caption{\label{fig:ft_anisotropy} The anisotropy of the contacts in the system, i.e. the difference between the eigenvalues of the fabric tensor $\mathbf{R}$.}
\end{figure}

In Fig.~\ref{fig:ft_anisotropy} we plot $\Delta\lambda$ vs $\phi$, for several different strain rates $\dot\gamma$.  We see that $\Delta\lambda$ peaks around $\phi\approx 0.3$ and then rises again as one crosses above jamming. This behavior follows the same trend as the contribution to the rotation from $\tau^\mathrm{el}$, as seen by looking at the difference between the curves of $\bar\omega/\dot{\gamma}$ and $\langle f(\theta_i) \rangle$ in Fig.~\ref{fig:angular_velocity}.  Thus, as might be expected, the magnitude  of the contribution of $\tau^\mathrm{el}$ to the rotation is related to the degree of anisotropy of contacts.  

We find that the eigenvector associated with eigenvalue $\lambda_1$ is oriented with an angle near $-30^\circ$ for all packing fractions, which means that the points of contact tend to be located in the upper-left or lower-right quadrants of the staples.
For a dilute system we can make sense of this by realizing that most new contacts will be made in these quadrants due to the shear flow $\mathbf{v}_\mathrm{av}(\mathbf{r})= y\dot{\gamma} \mathbf{\hat x}$.
While the torque generated by such a contact depends on the orientation of the staple, more often than not it will tend to increase rotation.
We note that near and above the jamming transition the anisotropy $\Delta\lambda$ depends on the shear rate $\dot\gamma$, while the rotation $\bar\omega/\dot\gamma$ does not.
Thus there may be some additional phenomenon due to the concave nature of the staples that is affecting the rotation.

\begin{figure}
\centering
\includegraphics[width=0.95\columnwidth]{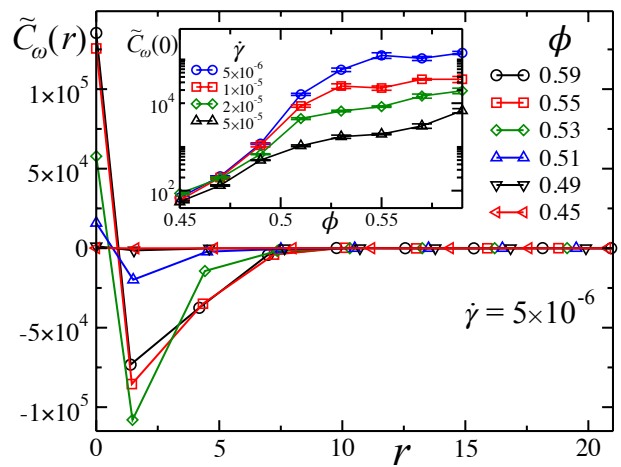}
\caption{\label{fig:Comega} Correlation of angular velocities, $\tilde C_\omega(r)\equiv[\langle \dot\theta(0)\dot\theta(r)\rangle -\langle\dot\theta\rangle^2]/\dot\gamma^2$, where we have averaged over all directions of the separation $\mathbf{r}$, vs $r=|\mathbf{r}|$, for several different packing fractions $\phi$.  The strain rate is fixed at $\dot\gamma=5\times 10^{-6}$.  The inset shows the variance of the angular velocity, $\tilde C_\omega(0)\equiv[\langle\dot\theta^2\rangle -\langle\dot\theta\rangle^2]/\dot\gamma^2$, vs $\phi$ for several different strain rates $\dot\gamma$.  For computing the correlation, we have used a bin width of $\Delta r =2D$, equal to twice the staple diameter, for measuring staple separations.
}
\end{figure}

Finally we consider the spatial correlations of the angular velocity, defining the correlation function,
\begin{equation}
\tilde C_\omega(\mathbf{r})\equiv\dfrac{\langle\dot\theta(0)\dot\theta(\mathbf{r})\rangle-\langle\dot\theta_i\rangle^2}{\dot\gamma^2},
\end{equation}
and $\tilde C_\omega(r)$ as the angular average of $\tilde C_\omega(\mathbf{r})$ over all orientations of the separation $\mathbf{r}$.
In Fig.~\ref{fig:Comega} we plot $\tilde C_\omega(r)$ vs $r$ for several different packing fractions $\phi$, at a fixed strain rate $\dot\gamma=5\times 10^{-6}$.  We see that $\tilde C_\omega(0)$ is positive, then rapidly drops negative on a length $r\approx 2$, before decaying to zero on the length scale of one staple $\sim 7$.  Thus, as was found for orientational order, there is essentially no spatial ordering of the angular velocity $\dot\theta$.  The negative value of $\tilde C_\omega(r)$ at short distances is due to the tendency of neighboring staples in contact to rotate in opposite directions, as would two meshed gears.

The variance of the angular velocity is given by,
\begin{equation}
\tilde C_\omega(0)=\dfrac{\langle \dot\theta_i^2\rangle -\langle\dot\theta_i\rangle^2}{\dot\gamma^2},
\end{equation}
which we plot vs $\phi$, for several different strain rates $\dot\gamma$, as the inset to Fig.~\ref{fig:Comega}.
We see that the fluctuations in $\dot\theta_i/\dot\gamma$ grow larger as the system gets denser, and as the strain rate decreases.

\section{Conclusions}

We have studied the jamming transition in a system of concave, frictionless, U-shaped particles under both compression and steady-state shearing.  For our particular particles, we have found that the jamming transition upon compressing is clearly associated with isostaticity, and occurs at values of $\phi_J^\mathrm{comp}$ between 0.49 and 0.5 depending on the initial packing fraction from which the compression begins.  In steady-state shearing our results suggested a lower bound for the jamming transition to be $\phi_J^\mathrm{shear}\ge0.52$.  
Several features we have observed, that (i) compression-driven jamming occurs at the isostatic point, that (ii) compression-driven jamming is influenced by the ensemble of initial states from which the compression begins, that (iii) shear-driven jamming occurs at a higher packing fraction than found from compressing dilute systems, and that (iv) there is a diverging translational correlation length as the shear driven jamming transition is approached, are all in common with the behavior observed for the jamming of frictionless disks.  
The main new effects  that we have observed are: (i) The time required to reach steady-state, independent of the initial configuration, in shear driven flow is very much longer than for disks.  (ii) The asymmetric shape of our particles leads to a non-uniform tumbling motion under shear flow, with corresponding nematic and tetratic orientational order; however, we find that the orientational ordering in general decreases as the jamming transition is approached from below, and the average angular velocity of the particles increases.  This is opposite to what we have observed in sheared frictionless rods \cite{DVH}, where orientational order increases and angular velocity decreases as density increases.  It is natural to attribute both of these effects (i) and (ii) to the concave particle shape of our  staples and the resulting geometric cohesion, however further work remains to be done to more firmly establish the effect of particle shape on orientational ordering and angular velocity.  Finally, we have also found (iii) that the nematic ordering appears be vanishing above jamming, while the tetratic ordering is growing.


Several works \cite{Buchholtz.PhysA1994,GalindoTorres.PRE2009,Papanikolaou.PRL2013} have suggested that geometric roughness on the surface of otherwise frictionless particles may provide a good model for the inter-particle tangential frictional forces that are usually present in dry granular systems.  Such roughness has been modeled \cite{Papanikolaou.PRL2013} by asperities on the surface of spherical particles, leading to a concave particle surface.  One may therefore ask if the concave staples studied in the present work display any of the features usually associated with the jamming of frictional particles. Our results, however, do not seem to find so.

In looking at Fig.~\ref{fig:compress}, we see that the apparent jamming transition $\phi_J$ moves to slightly {\em higher} values as the compression rate $\epsilon$ {\em decreases}.  
Similar results have been found for frictionless spheres \cite{Jodrey.PRA1985,Torquato.PRL2000}.  This is in contrast to what is observed for {\em frictional} spheres, where the {\em slower} the compression rate the {\em lower} $\phi_J$ one finds \cite{Onoda.PRL1990,Makse.Nature2008,Menon.SoftMatter2010}.  
Numerical simulations of slowly sheared frictional systems  \cite{Otsuki.PRE2011} show a discontinuous jump in the pressure at jamming, provided the friction coefficient is not too small.  Our results in Fig.~\ref{fig:shear_pz}a do not give any sign of such a discontinuous jump.  
We thus conclude that, for our staple shaped particles, inter-particle friction and geometric cohesion likely play quite different roles in the phenomenological behavior of granular materials. It remains to be seen if this conclusion holds more generally true for other particle shapes.

\section*{Appendix}

The function $f(\theta)$ that appears in Eq.~(\ref{eq:shear}) for the dissipative torque on a sheared staple is,
\begin{equation}
f(\theta) \equiv \dfrac{\displaystyle\int\limits_\mathrm{staple} d\mathbf{r}\,y^2}{\displaystyle\int\limits_\mathrm{staple} d\mathbf{r}\,|\mathbf{r}|^2},
\end{equation}
where $\mathbf{r}$ measures the distance from the staple's center of mass, and $\theta$ is the angle that the staple's spine makes with the $\mathbf{\hat x}$ axis.
We will approximate this integration by treating the staple as three connected, infinitesimally thin, rods with spine having a length $w$ and barbs each having length $\ell$, otherwise in the same arrangement as shown in Fig.~\ref{fig:staple}.  

\begin{figure}
\centering
\includegraphics[width=0.6\columnwidth]{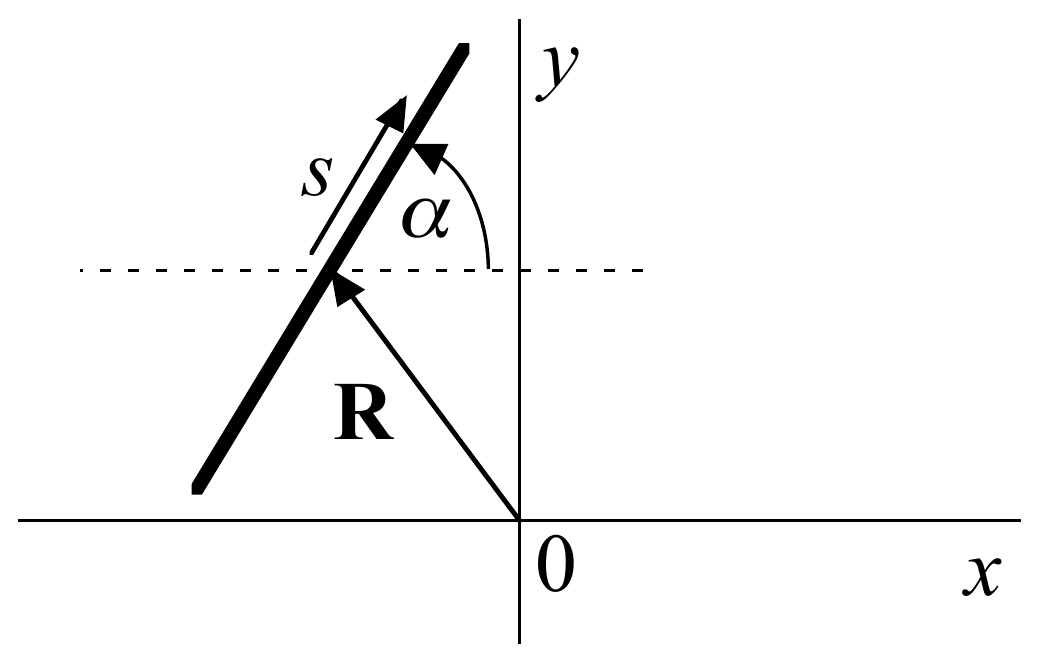}
\caption{\label{fA1}Geometry of a rod.
}
\end{figure}

We consider first the general case of a rod of length $L$, centered at a position $\mathbf{R}=(X,Y)$ and oriented at an angle $\alpha$ with respect to the $\mathbf{\hat x}$ axis, as shown in Fig.~\ref{fA1}. If $s$ is a coordinate that runs down the length of the rod from $-L/2$ to $L/2$, we then have,
\begin{equation}
\begin{array}{rl}
\displaystyle\int\limits_\mathrm{rod}d\mathbf{r}\,y^2 &= \displaystyle\int\limits_{-\frac{L}{2}}^{\frac{L}{2}}ds \left(Y+s\sin\alpha\right)^2
\\[12pt]
&=Y^2L+\dfrac{L^3}{12}\sin^2\alpha
\end{array}
\label{eA1}
\end{equation}
while
\begin{equation}
\begin{array}{rl}
\displaystyle\int\limits_\mathrm{rod}d\mathbf{r}\,|\mathbf{r}|^2 &= \displaystyle\int\limits_{-\frac{L}{2}}^{\frac{L}{2}}ds 
\left[(X+s\cos\alpha)^2+(Y+s\sin\alpha)^2\right]
\\[12pt]
&=|\mathbf{R}|^2L+\dfrac{L^3}{12}
\end{array}
\label{eA2}
\end{equation}

\begin{figure}
\centering
\includegraphics[width=0.6\columnwidth]{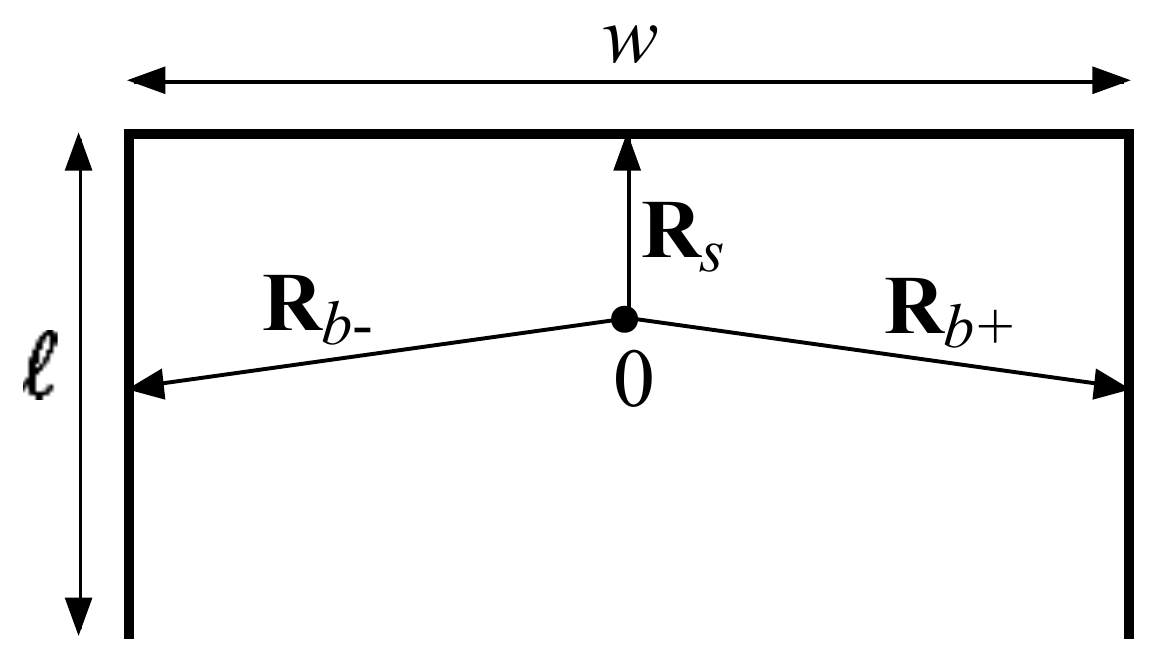}
\caption{\label{fA2}Geometry of a staple.
}
\end{figure}

To apply this to our staple, we consider first the situation when the staple is oriented at $\theta=0$, with the spine parallel to the $\mathbf{\hat x}$ axis and the barbs in the negative $\mathbf{\hat y}$ direction, as shown in Fig.~\ref{fA2}.  
If we set the origin of our coordinates at the center of mass of the three rods comprising the staple, then the spine is centered at position $\mathbf{R}_s=c\mathbf{\hat y}$, where
$c \equiv \ell^2/(w + 2 \ell)$.  The barbs are centered at positions $\mathbf{R}_{b\pm}=\pm(w/2) \mathbf{\hat x} + (c - \ell/2) \mathbf{\hat y}$.  Hence we have $|\mathbf{R}_s|^2=c^2$ and $|\mathbf{R}_{b\pm}|^2=w^2/4+(c-\ell/2)^2$.  When the staple is rotated through an angle $\theta$, we have for the resulting $Y$-components, $Y_s=c\cos\theta$, $Y_{b\pm}=\pm (w/2)\sin\theta+(c-\ell/2)\cos\theta$.

We now apply Eqs.~(\ref{eA1}-\ref{eA2}) to each segment of our staple, using for the spine $L_s=w$, $\alpha_s=\theta$, and for the barbs $L_b=\ell$, $\alpha_b=\theta+\pi/2$, and the above values of $|\mathbf{R}_s|^2$, $|\mathbf{R}_{b\pm}|^2$, $Y_s$ and $Y_{b\pm}$.  Adding the results we get,
\begin{equation}
\displaystyle\int\limits_\mathrm{staple}d\mathbf{r}\,y^2=\left(\dfrac{w^3}{12}+\dfrac{w^2\ell}{2}\right)\sin^2\theta
+\left(\dfrac{2\ell^3}{3}-c\ell^3\right)\cos^2\theta
\label{eq:inty2}
\end{equation}
and
\begin{equation}
\displaystyle\int\limits_\mathrm{staple}d\mathbf{r}\,|\mathbf{r}|^2=
\dfrac{w^3}{12}+\dfrac{w^2\ell}{2}+\dfrac{2\ell^3}{3}-c\ell^2
\label{eq:intr2}
\end{equation}

Dividing Eq. (\ref{eq:inty2}) by (\ref{eq:intr2}) we find,
\begin{equation}
f(\theta) = k\sin^2\theta+(1-k)\cos^2\theta,
\end{equation}
where
\begin{equation}
k=\frac{w^3/12 + w^2 \ell/2}{w^3/12 + w^2 \ell/2 + 2 \ell^3/3 - c \ell^2}.
\end{equation}
Finally, substituting back in $c = \ell^2/(w + 2 \ell)$, we can simplify $k$ as a function of a single variable $b \equiv \ell / w$, the barb to spine ratio.  And defining $C \equiv 2k - 1$ we have
\begin{equation}
 f(\theta) = \frac{1- C \cos 2 \theta}{2}
 \end{equation}
where
\begin{equation}
C = \frac{1+ 8 b + 12 b^2 - 8 b^3 - 4 b^4}{1+ 8 b + 12 b^2 + 8 b^3 + 4 b^4}.
\end{equation}
We note that $C = 1$ when $b = 0$ and $C \rightarrow -1$ when $b \rightarrow \infty$.

\section*{Acknowledgements}

We wish to thank D. V{\aa}gberg and A. Loheac for their assistance in the early stages of this work.  This work was supported by NSF Grants No. CBET-1133126 and CBET-1133722.   Computations were carried out on the GPU cluster at the University of Rochester's Center for Integrated Research Computing.

\bibliographystyle{granulmat} 

\end{document}